\documentclass[dvips]{acta}
\usepackage{supertabular,lscape,epsfig}
\usepackage{amssymb}
\usepackage{amsmath}
\usepackage{placeins}
\DeclareSymbolFont{ppa}{OT1}{ppl}{m}{it}
\DeclareMathSymbol{\vv}{\mathalpha}{ppa}{'166}

\newfont{\hb}{rphvb at 10pt}
\newfont{\hbo}{rphvbo at 10pt}
\newfont{\bitt}{rptmbi at 12pt}
\newfont{\bits}{rptmbi at 11pt}

\SetPages{1}{1}

\SetVol{66}{2016}

\begin{document}

\newcommand{\TabApp}[2]{\begin{center}\parbox[t]{#1}{\centerline{
  {\bf Appendix}}
  \vskip2mm
  \centerline{\small {\spaceskip 2pt plus 1pt minus 1pt T a b l e}
  \refstepcounter{table}\thetable}
  \vskip2mm
  \centerline{\footnotesize #2}}
  \vskip3mm
\end{center}}

\newcommand{\TabCapp}[2]{\begin{center}\parbox[t]{#1}{\centerline{
  \small {\spaceskip 2pt plus 1pt minus 1pt T a b l e}
  \refstepcounter{table}\thetable}
  \vskip2mm
  \centerline{\footnotesize #2}}
  \vskip3mm
\end{center}}

\newcommand{\TTabCap}[3]{\begin{center}\parbox[t]{#1}{\centerline{
  \small {\spaceskip 2pt plus 1pt minus 1pt T a b l e}
  \refstepcounter{table}\thetable}
  \vskip2mm
  \centerline{\footnotesize #2}
  \centerline{\footnotesize #3}}
  \vskip1mm
\end{center}}

\newcommand{\MakeTableApp}[4]{\begin{table}[p]\TabApp{#2}{#3}
  \begin{center} \TableFont \begin{tabular}{#1} #4 
  \end{tabular}\end{center}\end{table}}

\newcommand{\MakeTableSepp}[4]{\begin{table}[p]\TabCapp{#2}{#3}
  \begin{center} \TableFont \begin{tabular}{#1} #4 
  \end{tabular}\end{center}\end{table}}

\newcommand{\MakeTableee}[4]{\begin{table}[htb]\TabCapp{#2}{#3}
  \begin{center} \TableFont \begin{tabular}{#1} #4
  \end{tabular}\end{center}\end{table}}

\newcommand{\MakeTablee}[5]{\begin{table}[htb]\TTabCap{#2}{#3}{#4}
  \begin{center} \TableFont \begin{tabular}{#1} #5 
  \end{tabular}\end{center}\end{table}}

\newfont{\bb}{ptmbi8t at 12pt}
\newfont{\bbb}{cmbxti10}
\newfont{\bbbb}{cmbxti10 at 9pt}
\newcommand{\uprule}{\rule{0pt}{2.5ex}}
\newcommand{\douprule}{\rule[-2ex]{0pt}{4.5ex}}
\newcommand{\dorule}{\rule[-2ex]{0pt}{2ex}}
\def\thefootnote{\fnsymbol{footnote}}
\begin{Titlepage}
\Title{The OGLE Collection of Variable Stars.\\
Over 450\,000 Eclipsing and Ellipsoidal Binary Systems\\
Toward the Galactic Bulge\footnote{Based on observations
obtained with the 1.3-m Warsaw telescope at the Las Campanas Observatory of
the Carnegie Institution for Science.}}
\Author{I.~~S~o~s~z~y~ñ~s~k~i$^1$,~~
M.~~P~a~w~l~a~k$^1$,~~
P.~~P~i~e~t~r~u~k~o~w~i~c~z$^1$,~~
A.~~U~d~a~l~s~k~i$^1$,\\
M.\,K.~~S~z~y~m~a~ñ~s~k~i$^1$,~~
£.~~W~y~r~z~y~k~o~w~s~k~i$^1$,~~
K.~~U~l~a~c~z~y~k$^{1,2}$,~~
R.~~P~o~l~e~s~k~i$^{1,3}$,\\
S.~~K~o~z~³~o~w~s~k~i$^1$,~~
D.\,M.~~S~k~o~w~r~o~n$^1$,~~
J.~~S~k~o~w~r~o~n$^1$,~~
P.~~M~r~ó~z$^1$\\
and~~A.~~H~a~m~a~n~o~w~i~c~z$^1$}
{$^1$Warsaw University Observatory, Al.~Ujazdowskie~4, 00-478~Warszawa, Poland\\
e-mail: soszynsk@astrouw.edu.pl\\
$^2$Department of Physics, University of Warwick, Gibbet Hill Road, Coventry, CV4~7AL,~UK\\
$^3$Department of Astronomy, Ohio State University, 140 W. 18th Ave., Columbus, OH~43210, USA}
\Received{December 22, 2016}
\end{Titlepage}

\Abstract{We present a collection of 450\,598 eclipsing and ellipsoidal
  binary systems detected in the OGLE fields toward the Galactic bulge. The
  collection consists of binary systems of all types: detached,
  semi-detached, and contact eclipsing binaries, RS~CVn stars, cataclysmic
  variables, HW~Vir binaries, double periodic variables, and even planetary
  transits. For all stars we provide the {\it I}- and {\it V}-band
  time-series photometry obtained during the OGLE-II, OGLE-III, and OGLE-IV
  surveys. We discuss methods used to identify binary systems in the OGLE
  data and present several objects of particular interest.}{binaries:
  eclipsing -- Catalogs}

\Section{Introduction}
In recent years, the number of known variable stars in the central regions
of the Milky Way has grown significantly, mostly thanks to the publication
of an extensive collection of variable stars by the Optical Gravitational
Lensing Experiment (OGLE). Hundreds of thousands of pulsating stars (\eg
Soszyñski \etal 2011b, 2013, 2014) discovered by OGLE have enabled thorough
analyses of, both, the stellar pulsation theory and the structure of the
Galactic bulge. Eclipsing binary systems have the potential to play an
equally important role in the exploration of the central parts of the
Galaxy, although until now only a small fraction of binaries in this region
of the sky has been cataloged and studied.

Eclipsing binary systems offer an opportunity to directly measure the
fundamental stellar parameters, such as masses, sizes, temperatures, absolute
luminosities, and rotation (Andersen 1991). Binary systems serve as
testbeds for stellar evolutionary theories, since both components have the
same age and chemical composition. Eclipsing binaries are used to study
dynamical interactions between stars, mass exchange and loss, stellar
magnetic activity, limb darkening, and tidal circularization
theories. Moreover, detached eclipsing binaries are accurate distance
indicators in the Milky Way and other galaxies (\eg Paczyñski 1997, Kaluzny
\etal 2013, Pietrzyñski \etal 2013).

First eclipsing binaries in the region of the Galactic bulge were
discovered at the turn of the twentieth century (Roberts 1895, Pickering
and Leavitt 1904, Pickering 1908). In subsequent years, thanks to the
efforts of many observers (\eg Parenago 1931, Swope 1938, Ferwerda 1943,
Baade 1946, Plaut 1948, 1958, 1971, Gaposchkin 1955, Kooreman 1966), the
number of known eclipsing binaries toward the central regions of the Milky
Way has grown to about 300.

At the end of the twentieth century with the advent of large-scale
variability surveys, this sample has increased significantly. About 1650
eclipsing binary systems have been identified by Udalski \etal (1994,
1995ab, 1996, 1997) in the photometric database obtained during the first
phase of the OGLE project. This sample was extended by 1575 contact binary
systems identified by Szymañski \etal (2001). Groenewegen (2005) used the
OGLE-II photometry of variable stars in the Galactic bulge (Wo¼niak \etal
2002) to select 3053 detached eclipsing binaries, mostly suited for
distance determinations. Devor (2005) used the same data to fit over 10\,000
models of eclipsing binaries and to identify 3170 detached systems. This
list should be supplemented by 59 systems with very shallow eclipses
published by Udalski \etal (2002ab). Two of these objects, OGLE-TR-10 and
OGLE-TR-56, turned out to be planetary systems (Konacki \etal 2003, 2005)
-- the first ones discovered with the transit method. Additionally, a few
hundred eclipsing and ellipsoidal binaries toward the center of the Milky
Way have been discovered by Pojmañski and Maciejewski (2004, 2005) based on
the observations gathered by the All Sky Automated Survey (ASAS).

In this paper, we present the OGLE collection of eclipsing and ellipsoidal
binary systems in the Galactic bulge. Our collection, consisting of more than
450\,000 objects, not only increases the number of known binary stars in
the central regions of the Milky Way by two orders of magnitude, but also
multiplies the total number of eclipsing binaries known to date in the whole
Universe. The first stars from our collection -- 242 ultra-short-period
binary systems ($P_{\rm orb}<0.22$~d) -- have already been published
by Soszyñski \etal (2015). One of these stars -- OGLE-BLG-ECL-000066 --
with an orbital period of 0.0984~d is probably the binary system consisting
of non-degenerate components with the shortest known orbital period. The
paper is structured as follows. In Section~2, we present the photometric
data used in the analysis. Section~3 describes the selection and
classification of the binary systems in the Galactic bulge. In Section~4,
we present the collection itself and estimate its completeness. In
Sections~5 and 6, we discuss and summarize our results.

\Section{Observations and Data Reduction}
Our collection of binary systems in the Galactic bulge is based on the
photometric data collected by the OGLE survey between 1997 and 2015 at Las
Campanas Observatory, Chile, with the 1.3-m Warsaw Telescope. The
observatory is operated by the Carnegie Institution for Science. In
1997--2000, during the OGLE-II stage, about 30 million stars in the area of
11~square degrees in the central parts of the Milky Way were constantly
monitored. In 2001, with the beginning of the OGLE-III survey, the sky
coverage was extended to nearly 69~square degrees and the number of
monitored stars increased to 200 million. Finally, from 2010 until today
the OGLE-IV project regularly observes about 400~million stars in 182
square degrees of the densest regions of the Galactic bulge. Our search for
eclipsing variables was based primarily on the OGLE-IV data.

The Warsaw telescope is currently equipped with a 32-detector mosaic CCD
camera covering an area of about 1.4~square degrees on the sky. Most of the
observations were made through the Cousins {\it I}-band filter with an
integration time of 100~s. The number of collected data points varies
greatly between individual fields, with the least-sampled fields having
only about 100 {\it I}-band observations per star while those most sampled
contain over 12\,000 observations. The CCD saturation limit is about 13~mag
in the {\it I}-band, while the faintest stars in the OGLE database have
$I\approx21$~mag. Typical photometric uncertainties of individual
measurements for bright stars are about 0.005~mag. Up to 10\% of the
observations were secured in the Johnson {\it V}-band with the exposure
time of 150~s.

The OGLE photometry was carried out using the Difference Image Analysis
technique (DIA, Alard and Lupton 1998, Wo¼niak 2000). The instrumental
photometry was calibrated to the standard system with the procedure
described by Udalski \etal (2015). The accuracy of the zero point of this
transformation is at the level of 0.02~mag, however the brightness of some
individual stars may be significantly affected by blending, crowding,
reflections from bright stars, etc. More details on the OGLE
instrumentation, photometric reductions and astrometric calibrations are
provided by Udalski \etal (2015).

\Section{Selection and Classification of Binary Systems}
The selection of eclipsing and ellipsoidal binary systems from the set of
400 million stars requires the use of automated mechanisms of the variable
star classification. However, each light curve included in our collection
has been visually inspected by a human at least once, so our procedure can
be described as semi-automatic.

In order to detect binary systems, we performed an extensive period search
for all stars observed by OGLE toward the Galactic bulge. We applied two
different methods of the period search to every {\it I}-band light curve
stored in the OGLE database. The {\sc Fnpeaks}
code\footnote{\it http://helas.astro.uni.wroc.pl/deliverables.php?lang=en\&active=fnpeaks}
computes the Fourier amplitude spectra and provides the most significant
periods with their signal-to-noise ratios. The Fourier analysis is better
suited to detect periodicities in the magnitude sequences with continuous
light variations, \ie in contact, semi-detached, and ellipsoidal binary
systems. On the other hand, the Fourier techniques often fail in the case
of detached eclipsing binaries, in particular those with narrow
eclipses. Therefore, we additionally applied the Box-Least Squares (BLS)
period-search algorithm (Kov{\'a}cs \etal 2002) implemented in the {\sc
Vartools}
program\footnote{\it http://www.astro.princeton.edu/{\textasciitilde{}}jhartman/vartools.html}
(Hartman and Bakos 2016).

The preselection of candidates for binary systems was made with two
automatic methods. The first one used the machine-learning technique based
on the Random Forest algorithm (Breiman 2001). The details of this method
can be found in Pawlak \etal (2016). The second method used fitting of
the template light curves. As the templates we used {\it I}-band light
curves of bright ($I<17$~mag) eclipsing and ellipsoidal variables from
three best-sampled OGLE-IV fields: BLG501, BLG505, and BLG512. Each light
curve from these fields consists of at least 10\,000 data points. The
Julian Dates of the individual measurements were transformed to the orbital
phases and averaged in 1000 bins. The magnitudes were normalized in such a
way that the maximum brightness of every template was zero and the amplitude
was equal to~1. Fig.~1 presents several template light curves from our
set. The full collection consists of 747 templates.
\begin{figure}[p]
\hglue-3mm
{\includegraphics[width=13.1cm]{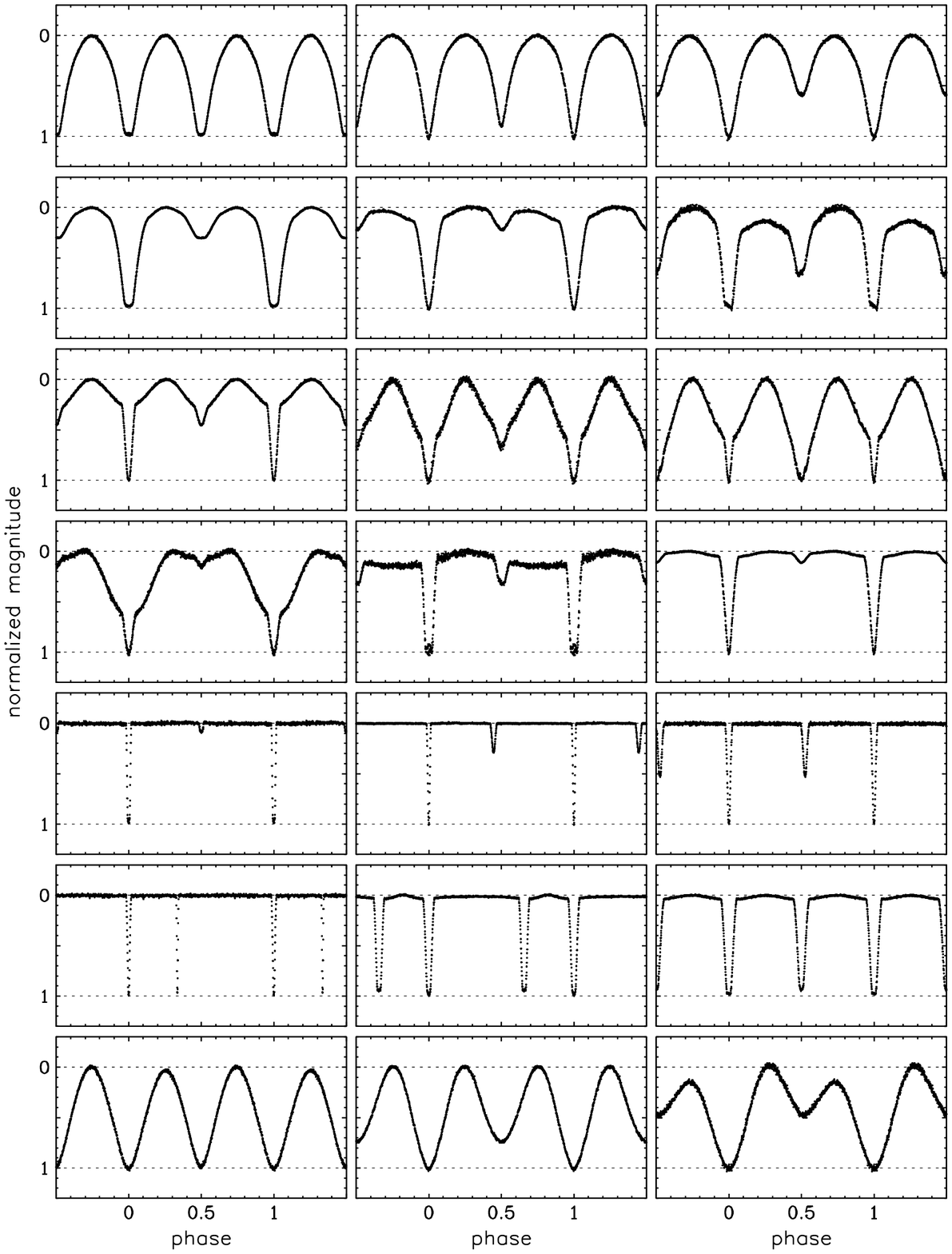}} 
\vspace*{4pt}
  \FigCap{Examples of template light curves of eclipsing and ellipsoidal
    binary systems. The templates were obtained from the {\it I}-band light
    curves of bright variables detected in three best-sampled OGLE-IV
    fields: BLG501, BLG505, and BLG512.}
\end{figure}

In the last step of the selection procedure, we visually inspected the
light curves of the best candidates for binary stars released by the
aforementioned automatic methods. Obvious false positives were removed from
the sample. We also rejected light curves that mimic eclipsing or
ellipsoidal variability, but probably have another origin. These are for
example non-eclipsing spotted variables (RS~CVn, BY~Dra, and Ap stars) or
red giant stars exhibiting the long secondary periods, although there are
strong arguments that this enigmatic phenomenon is related to binarity (\eg
Soszyñski and Udalski 2014). The final collection was supplemented by
eclipsing and ellipsoidal binaries identified during previous searches for
other types of variable stars, \eg Cepheids, RR~Lyr stars, or long-period
variables (\eg Soszyñ\-ski \etal 2011b, 2013, 2014).

The classification of our stars was mainly based on the light curve
template fitting. We divided the sample into three groups: candidates for
eclipsing contact binary systems (the status of these stars has to be
confirmed spectroscopically), non-contact eclipsing systems, and
ellipsoidal variables. Additionally, a special designation (CV) was
assigned to eclipsing cataclysmic variables -- post-novae and dwarf novae
-- recently cataloged by Mróz \etal (2015ab).

\Section{Binary Systems in the Galactic Bulge}
The OGLE collection of eclipsing and ellipsoidal binary systems toward the
Galactic bulge contains 450\,598 objects, of which 86\,560 stars are
candidates for contact systems, 338\,633 are probable semi-detached and
detached binaries (including 18 cataclysmic variables), and 25\,405 are
non-eclipsing ellipsoidal variables. The data on all these objects are
available through the OGLE anonymous FTP sites or via the OGLE web
interface:
\begin{center}
{\it ftp://ftp.astrouw.edu.pl/ogle/ogle4/OCVS/blg/ecl/}\\
{\it http://ogle.astrouw.edu.pl}
\end{center}

Each star has a unique identifier which follows the scheme introduced by
So\-szyñski \etal (2015). The identifiers OGLE-BLG-ECL-NNNNNN and
OGLE-BLG-ELL-NNNNNN (where NNNNNN is a six-digit number) have been given to
the eclipsing and ellipsoidal binaries, respectively. The stars are
arranged in order of increasing right ascension, with the exception of the
242 ultra-short-period binaries already published by Soszyñski \etal
(2015). For each object we provide its identifier, J2000 equatorial
coordinates, type of variability, {\it I}- and {\it V}-band magnitudes at
maximum light, orbital period, primary and secondary eclipse depths in the
{\it I}-band, and epoch of the primary eclipse minimum.

We also provide the time-series {\it I}- and {\it V}-band photometry
collected during the OGLE-II, OGLE-III, and OGLE-IV projects (if
available). The light curves from each stage of the survey were
independently calibrated to the standard Johnson-Cousins photometric
system, however smaller or bigger offsets between the photometric zero
points may occur for individual stars. This may be a result of different
instrumental configurations, in particular different filters and CCD
detectors, used in the three stages of the OGLE project, but also of
crowding and blending by unresolved stars which may randomly affect the
reference zero point of the DIA photometry. These offsets should be taken
into account when merging the light curves from different stages of the
project.

The periods and other observational parameters were derived using solely
the OGLE-IV light curves collected in the years 2010--2015 or, if the
OGLE-IV light curves were unavailable, using the OGLE-II or OGLE-III light
curves. The periods were refined using two programs: for the detached
systems we applied the BLS algorithm implemented in the {\sc Vartools}
program (Hartman and Bakos 2016), while in the remaining cases the periods
were derived with the {\sc Tatry} code based on the multi-harmonic
periodogram (Schwarzenberg-Czerny 1996). One should be warned that only
some of the folded light curves have been examined by eye after the final
determination of the periods and it cannot be excluded that some of the
periods provided in our collection are not real, for example they can be
two times longer or shorter than actual orbital periods of the systems. The
amplitudes and the luminosities at maximum have been derived fully
automatically with the template fitting.
\begin{figure}[htb]
\includegraphics[width=12.7cm]{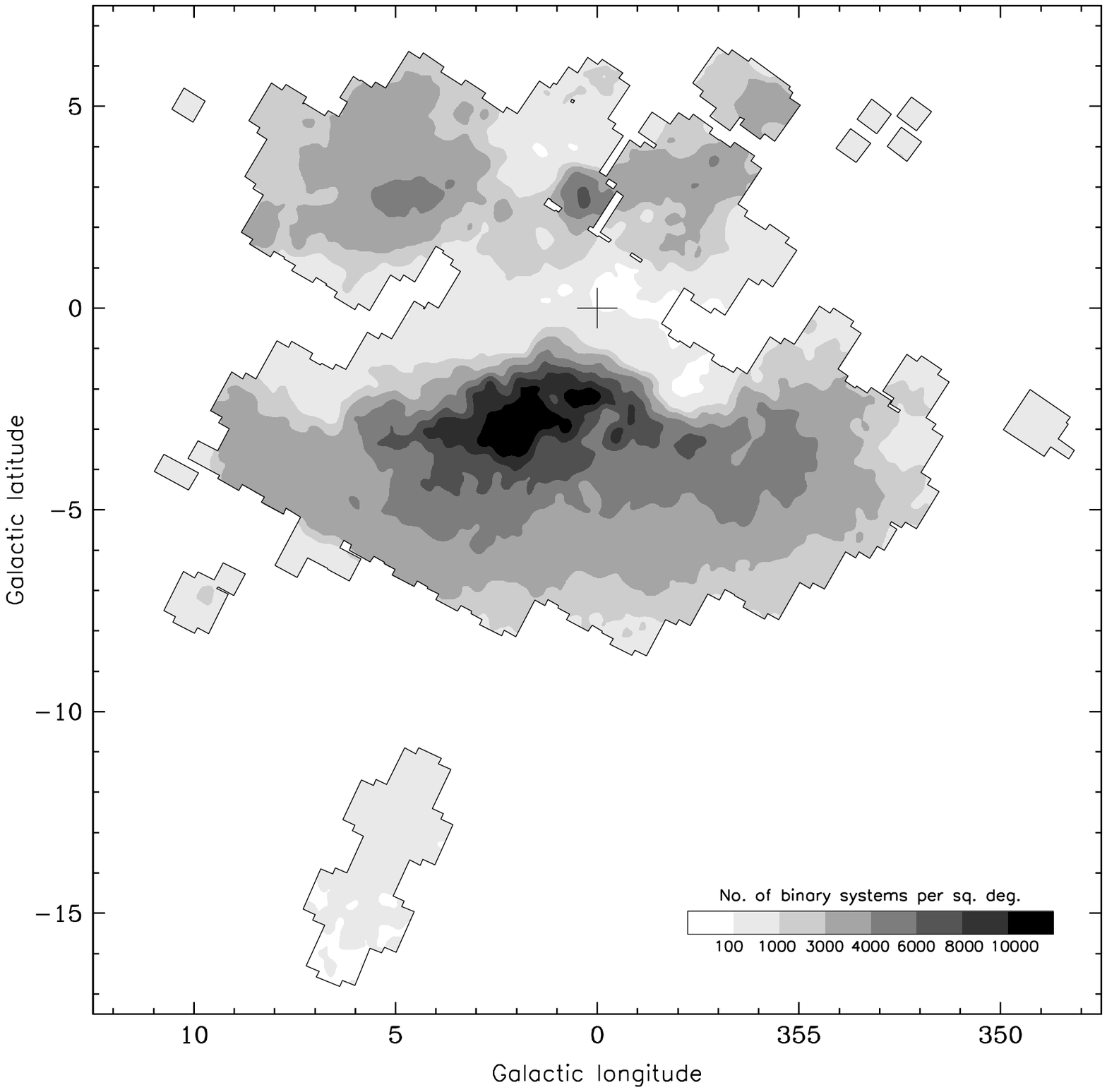}
\FigCap{Spatial distribution of eclipsing and ellipsoidal binary systems in
the OGLE fields toward the Galactic bulge.}
\end{figure}

In Fig.~2, we present the surface density map of the binary systems from
the OGLE collection. As one might have expected, the highest density is
observed around the Galactic coordinates $(l,b)=(2\arcd, -3\arcd)$, since
these are the densest and the most frequently observed OGLE fields. The
number of objects in our sample drastically decreases near the Galactic
plane due to large interstellar extinction in this region. All systems
observed here are likely located in the foreground of the Galactic bulge.

We estimated the completeness of our sample using stars located in the
overlapping parts of adjacent OGLE-IV fields. Such stars have double
entries in the OGLE database, since they were recorded twice, independently
in both fields. Note, that in the final version of our collection each star
is represented by a single OGLE-IV light curve, usually the one with larger
number of data points.

We found {\it a posteriori} that 10\,118 binaries in our collection have
such double entries in the OGLE database (assuming that both light curves
must consist of at least 100~points), so we had a chance to detect 20\,236
counterparts. We independently identified 16\,174 of them, which implies
the completeness of the whole sample at the level of 75\%. Of course, the
completeness is a strong function of the number of data points obtained for
a given star, brightness, amplitudes, and light curve morphology. For
example, the completeness for binary systems brighter than $I=18$~mag is
about 83\%, while for fainter stars it drops to 64\%. Obviously, these
estimates relate only to the binary systems which brightness, amplitudes
and light curve shapes are sufficient to be potentially identified and
classified with the OGLE photometry.  \vspace*{-9pt}
\Section{Discussion}
\vspace*{-5pt}
Our collection contains eclipsing and ellipsoidal binary stars of all
types: detached, semi-detached, and contact eclipsing binaries, ellipsoidal
variables, eclipsing RS~CVn stars (exhibiting additional variations due to
stellar spots), cataclysmic variables, HW~Vir binaries (consisting of a
cool main-sequence star and a B-type subdwarf), double periodic variables
(Mennickent \etal 2003), and even planetary transits (Udalski \etal
2002ab). The orbital periods of the systems range from 75~minutes (0.05~d)
to over 7~years (2600~d). For about twenty eclipsing binaries we cannot
currently determine the periods, since these stars exhibited only one or
two eclipses during the whole time span of the OGLE survey. Their periods
are probably longer than 8--20 years (depending on whether the star was
monitored during the OGLE-IV project only or also during the previous
stages of the survey). Detected variability amplitudes range from
milimagnitudes to several magnitudes.

The list of potential astrophysical application of our collection is
long. Time-series, standard OGLE photometry together with multi-epoch
spectroscopic observations may provide a complete set of parameters of both
components: their radii, masses, and effective temperatures. Our sample
contains many detached eclipsing binaries with deep eclipses which are
ideal for distance determinations, so they can be used to trace the
structure of the bulge. In turn, statistical properties of the collection
provide a useful window into the history of formation and evolution of
binary systems in the central regions of our Galaxy.

\begin{figure}[p]
\includegraphics[width=12.7cm]{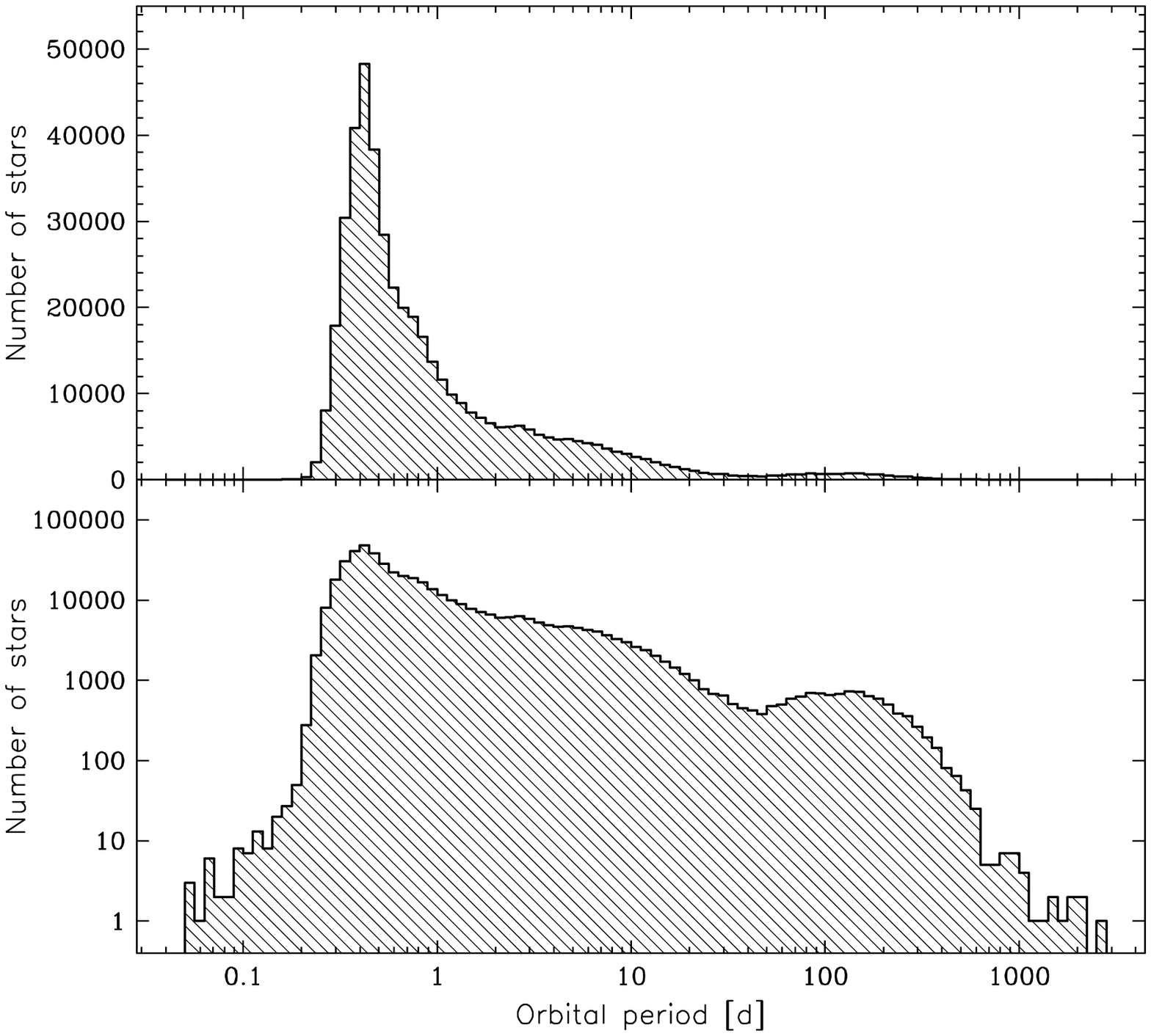}
\vskip3pt
\FigCap{Distribution of orbital periods of the OGLE collection of binary
systems in the Galactic bulge. The number of stars is presented in the
linear ({\it upper panel}) and logarithmic scale ({\it lower panel}).}
\end{figure}
\begin{figure}[p]
\includegraphics[width=12.7cm]{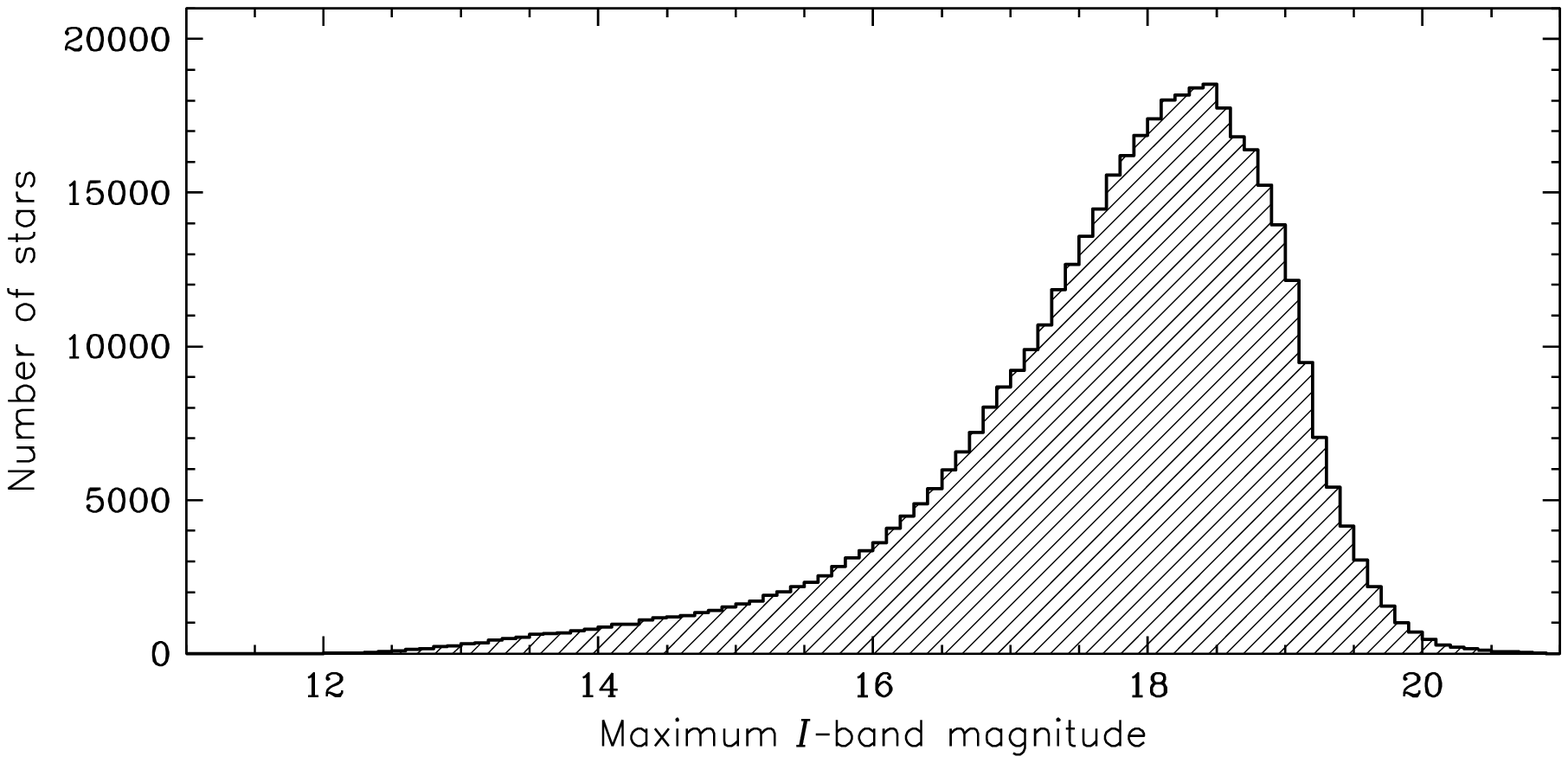}
\vskip3pt
\FigCap{Distribution of out-of-eclipse {\it I}-band magnitudes of the OGLE
collection of binary systems in the Galactic bulge.}
\end{figure}

Fig.~3 presents the distribution of orbital periods of the entire
sample. The number of objects shown in the linear scale (the upper panel of
Fig.~3) illustrates the proportions of various periods, while the
logarithmic scale (the lower panel of Fig.~3) is better suited to show
details of the distribution. The bulk of our systems have orbital periods
below 1~day and most of them are close binary systems consisting of
main-sequence stars. The period distribution has a strong maximum at about
0.40~d. A well-known sharp cut-off at a lower limit around 0.22~d is also
visible. The long-period binaries are dominated by very detached binaries or
close systems (usually ellipsoidal variables) consisting of red giant
stars. Such long-period ellipsoidal variables are responsible for a local
flat maximum visible in the distribution for periods between 100~d and
200~d. Red giants in close binary systems form period-luminosity relations
(\eg Pawlak \etal 2014). The distribution of apparent {\it I}-band
magnitudes at maximum light is shown in Fig.~4. The distribution peaks at
18.4~mag, but it is clear that the completeness of our collection begins to
decrease at $I=18$~mag. It is in agreement with our tests of completeness
presented in Section~4.

\begin{figure}[htb]
\includegraphics[width=12.7cm]{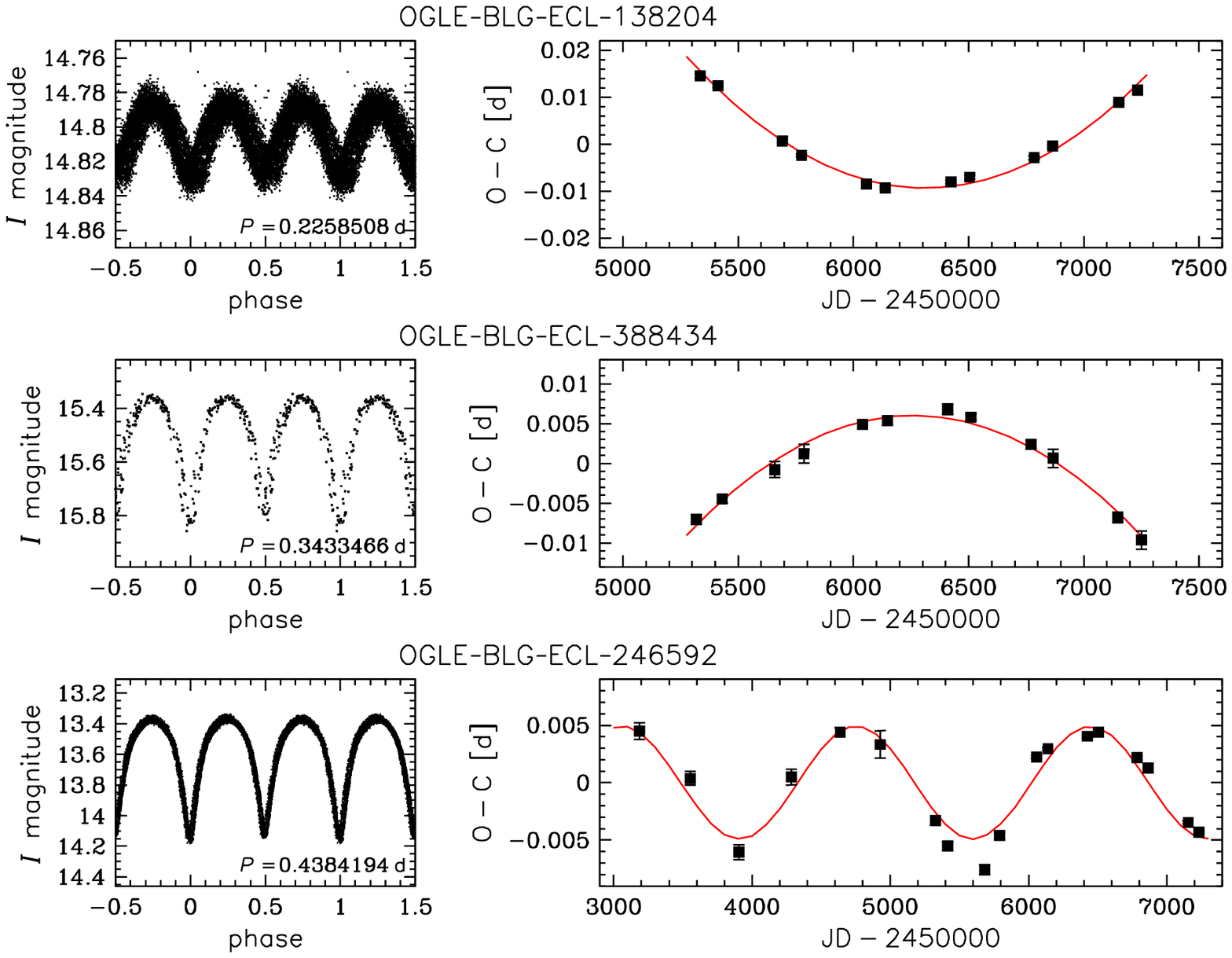}
\vskip1pt
\FigCap{Examples of three close binary systems with prominent period
changes. {\it Left panels} show light curves folded with the best constant
periods. {\it Right panels} present $O-C$ diagrams obtained for these
stars.}
\end{figure}
The long-term OGLE photometry can be used to test the stability of the
orbital periods of individual systems. Binary stars may change their
periods due to a mass transfer between the components, mass ejections from
the system, or a presence of an unseen tertiary companion. In Fig.~5, we
present three examples of close binary systems with prominent period
changes: monotonically increasing, decreasing, and cyclically varying
periods. A careful analysis of the entire sample should shed light on the
problems of short-period cut-off in contact binary systems, the role of
magnetic breaking in the evolution of close systems, and the abundance of
triple and other multiple systems.

\begin{figure}[b]
\centerline{\includegraphics[width=10.0cm]{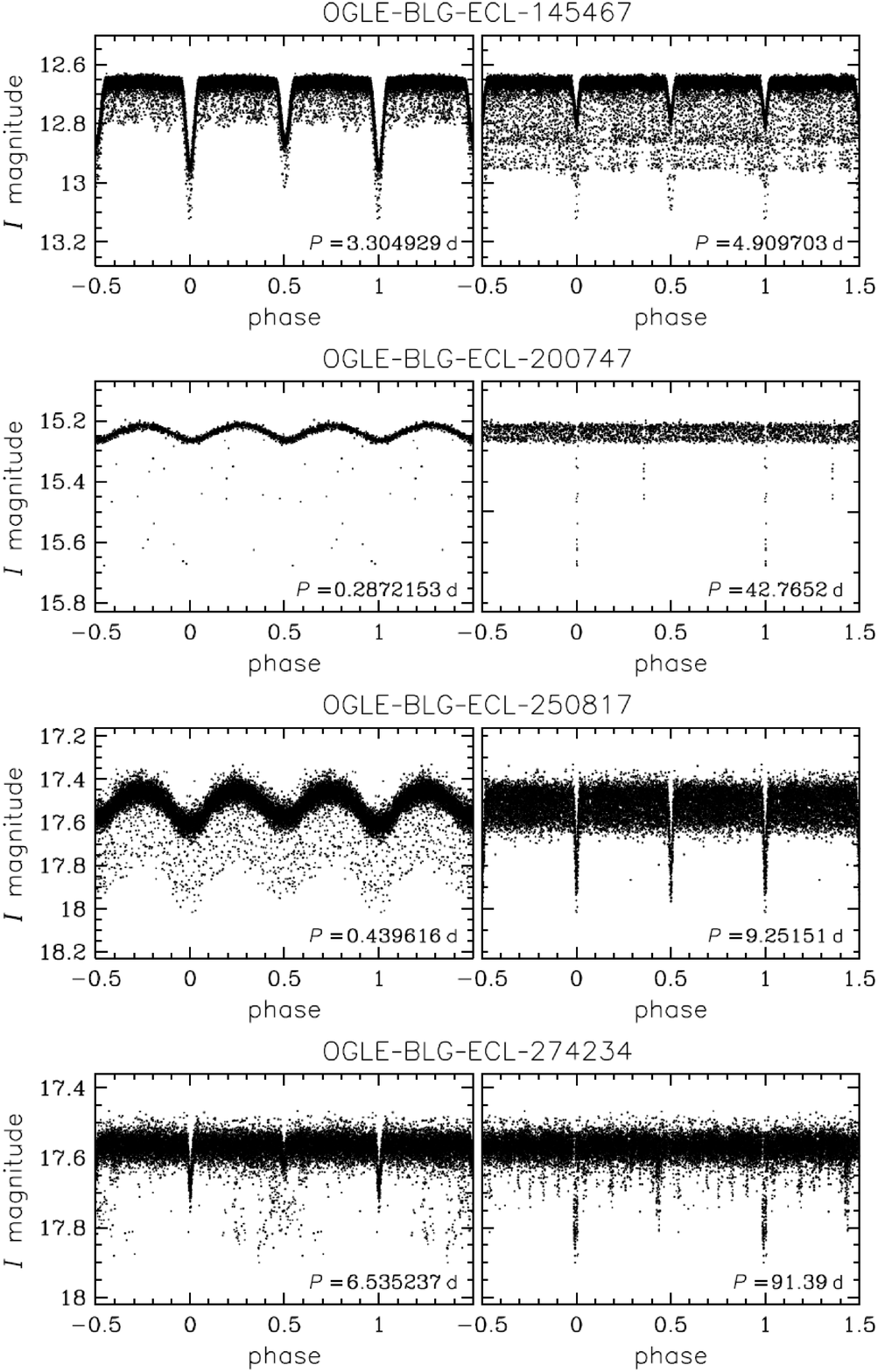}}
\vspace*{1pt}
\FigCap{Examples of four double binaries. Each pair of {\it left} and {\it
right panels} show the same light curve folded with different periods.}
\end{figure}
The latter issue can also be examined using double binaries -- objects in
which two superimposed eclipsing or ellipsoidal modulations are
simultaneously visible. Our sample contains at least several dozen such
stars. Four light curves of this type are presented in Fig.~6. Double
binaries can be blended stars observed on the same line of sight or can be
physically bound triple or quadruple systems. Significant changes of the
orbital period observed for example in OGLE-BLG-ECL-274234 (lower panel of
Fig.~6) suggest that in this case we deal with an interacting multiple
system. Binaries that ceased or began their eclipsing variations can also
be used to study triple systems. Graczyk \etal (2011), who discovered such
stars in the Large Magellanic Cloud, called them transient eclipsing
binaries. Three light curves of transient eclipsing binaries in the
Galactic bulge are shown in Fig.~7.
\begin{figure}[htb]
\includegraphics[width=12.7cm]{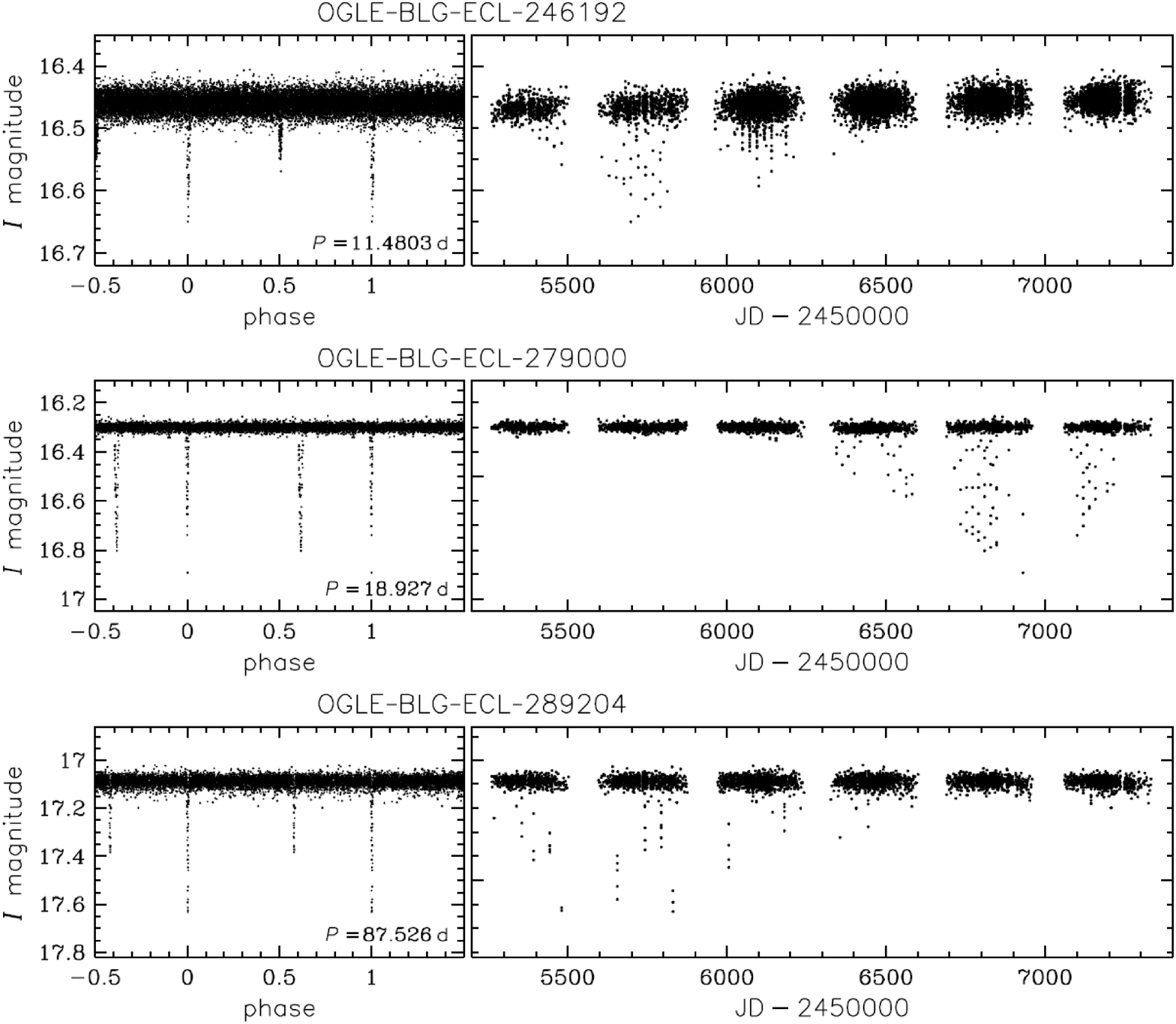}
\vskip1pt
\FigCap{OGLE-IV {\it I}-band light curves of three eclipsing binaries that
ceased or began their eclipsing variations due to precession of the orbital
planes. {\it Left panels} show light curves folded with the orbital
periods. {\it Right panels} show the same, unfolded light curves.}
\end{figure}

Eclipsing or ellipsoidal binaries that exhibit superimposed additional
types of variability deserve special attention, since such stars may
provide solutions to many astrophysical problems. For example,
OGLE-BLG-RRLYR-02792 -- a variable initially classified by Soszyñski \etal
(2011a) as an RR~Lyr star in an eclipsing binary system -- turned out to be
a first representative of a new type of pulsating stars -- so called binary
evolution pulsators (Pietrzyñski \etal 2012). OGLE-BLG-RRLYR-02792, as an
eclipsing binary, is also included in the present collection and designated
as OGLE-BLG-ECL-108266.

Our collection contains many other binary systems with additional
modulation of light. These objects are flagged in the remarks of the
collection. Fig.~8 shows light curves of four sample stars of this
type. The collection includes at least several eclipsing variables with
$\delta$~Sct-like variations, one more star classified as an RR Lyrae star
(OGLE-BLG-ECL-172630 = OGLE-BLG-RRLYR-06807), many long-period variables
(pulsating red giants), and several microlensing events. Probably, some of
these objects are blends with unresolved variables.
\begin{figure}[b]
\includegraphics[width=12.7cm]{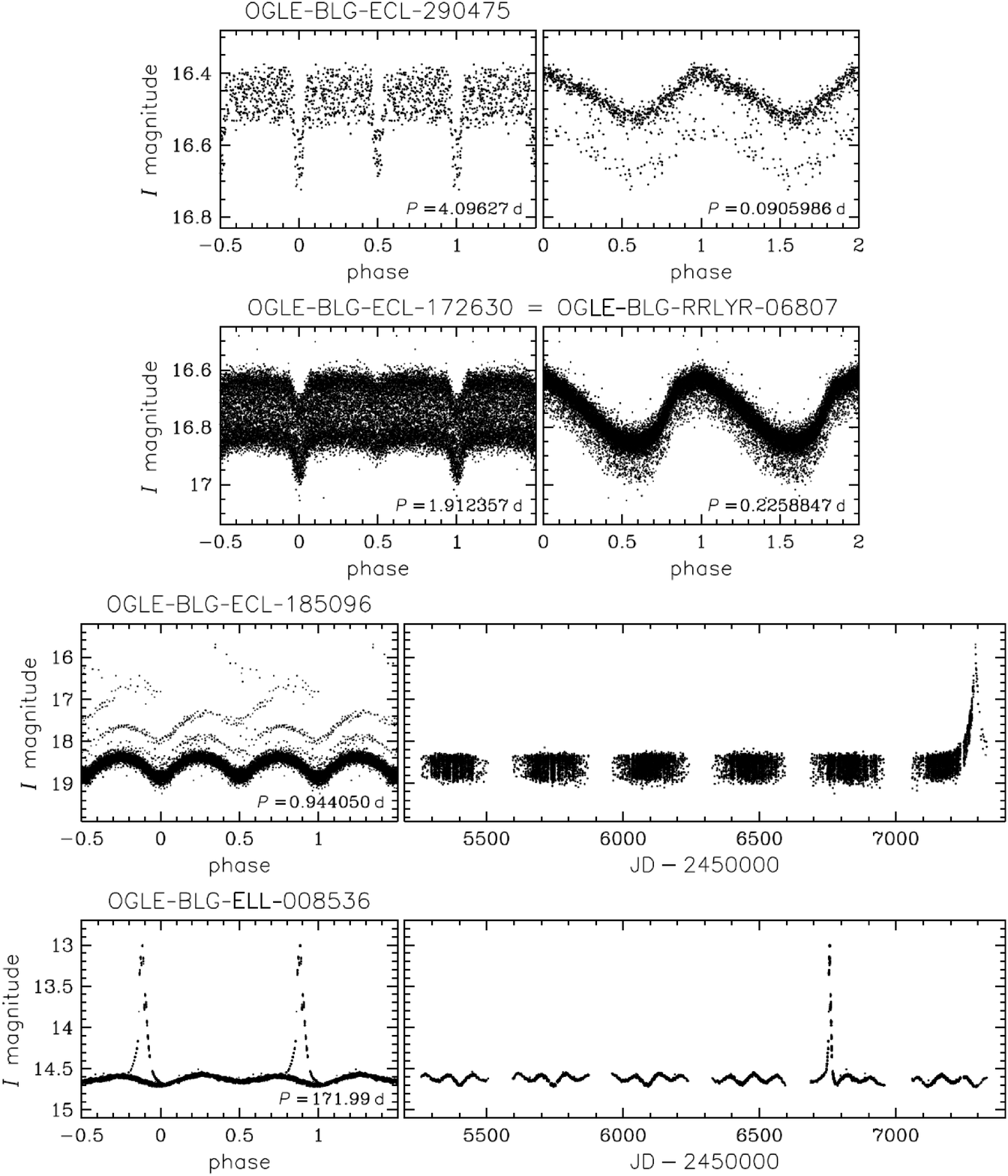}
\vskip5pt
\FigCap{OGLE-IV light curves of four eclipsing or ellipsoidal binaries with
a superimposed additional variability. Each pair of panels show the same
light curves. Two {\it upper panels} show pulsating stars ($\delta$~Sct and
RR~Lyr stars), while {\it lower panels} display microlensing events.}
\end{figure}

\begin{figure}[htb]
\includegraphics[width=12.3cm]{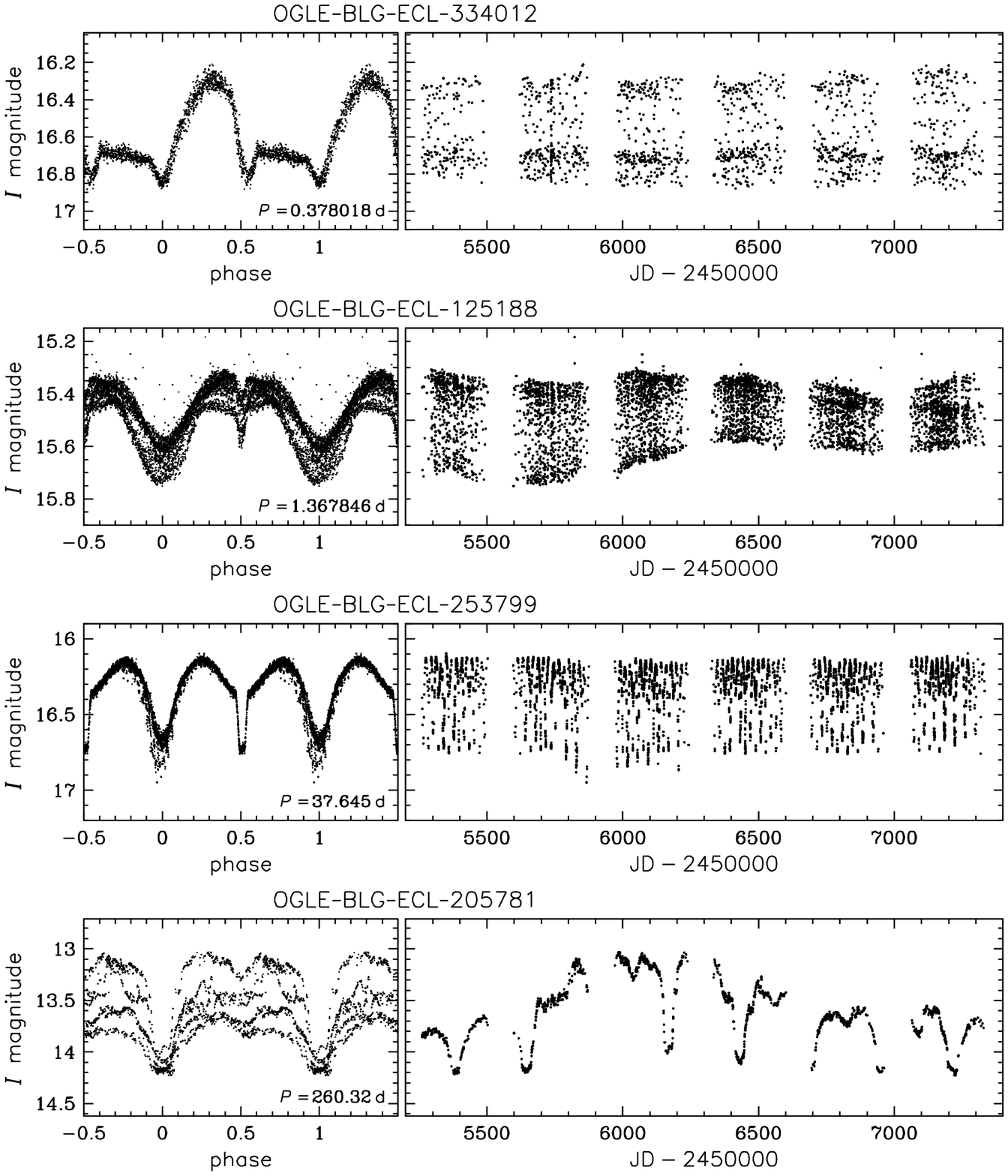}
\vskip1pt
\FigCap{Four examples of eclipsing binary systems with peculiar light
curves. {\it Left panels} show light curves folded with the orbital
periods. {\it Right panels} show the same, unfolded light curves.}
\end{figure}
Hundreds of eclipsing RS~CVn stars -- chromospherically active binary
systems -- have been included in the OGLE collection. Their light curves
usually exhibit two periodicities: the orbital period (manifested by the
eclipses) and the brightness fluctuation caused by the rotation of a
spotted star. The amplitudes and shapes of the out-of-eclipse light curves
gradually change from cycle to cycle due to changes of the spot
distribution on the stellar surface. Usually, both periods are similar due
to the synchronization of the orbital and rotation cycles, but there are
also RS~CVn stars in our collection which show very different
periodicities. For all RS~CVn stars we provide the orbital periods of the
systems, even if the variations due to spot surface coverage have larger
amplitudes than the amplitudes of eclipses.

\Section{Conclusions}
We presented the OGLE collection of eclipsing and ellipsoidal binary
systems in the Galactic bulge. Our sample multiplies the number of known
binary systems in this environment and in all other stellar
environments. Together with the collection of binary systems in the
Magellanic Clouds (Pawlak \etal 2016) and in the Galactic disk
(Pietrukowicz \etal 2013) the OGLE Collection of Variable Stars contains
about half a million binary systems. For each object we provide the
long-term time-series OGLE photometry in the {\it I} and {\it V} standard
photometric bands, well suited for studying properties of the individual
systems and the stellar environment in which they are located.

Such a huge collection of binary stars contains objects of particular
interest: systems with secular and cyclic period changes, double binaries,
transient eclipsing binaries, other types of variable stars in the
eclipsing configuration, systems with accretion disks, and many other
peculiar eclipsing and ellipsoidal variables. In Fig.~9, we present four
such unusual light curves. In the future, we plan to extend the OGLE
collection of the newly found systems in the same bulge fields and in the
additional fields covering practically the entire Galactic bulge with its
far outskirts. We also expect to discover hundreds of thousands of binary
systems in the Galactic disk, which is currently extensively observed by
the OGLE-IV Galactic Variability Survey.

\Acknow{We would like to thank Profs. M.~Kubiak and G.~Pietrzyñ\-ski,
former members of the OGLE team, for their contribution to the collection
of the OGLE photometric data over the past years. We are grateful to
Z.~Ko³aczkowski and A.~Schwar\-zen\-berg-Czerny for providing software used
in this study.

This work has been supported by the Polish National Science Centre grant
no. DEC-2011/03/B/ST9/02573. We gratefully acknowledge financial support
from the Polish Ministry of Science and Higher Education through the
program ``Ideas Plus'' award No. IdP2012 000162. MP acknowledges support
from the Polish National Science Centre grant PRELUDIUM
no. 2014/13/N/ST9/00075. The OGLE project has received funding from the
Polish National Science Centre grant MAESTRO no. 2014/14/A/ST9/00121.}


\begin{references}
\refitem{Alard, C., and Lupton, R.H.}{1998}{\ApJ}{503}{325}
\refitem{Andersen, J.}{1991}{Astronomy and Astrophysics Review}{3}{91}
\refitem{Baade, W.}{1946}{\PASP}{58}{249}
\refitem{Breiman, L.}{2001}{Machine Learning}{45}{5}
\refitem{Devor, J.}{2005}{\ApJ}{628}{411}
\refitem{Ferwerda, J.G.}{1943}{Bull. Astron. Inst. Netherlands}{7}{337}
\refitem{Gaposchkin, S.}{1955}{Peremennye Zvezdy}{10}{337}
\refitem{Graczyk, D., \etal}{2011}{\Acta}{61}{103}
\refitem{Groenewegen, M.A.T.}{2005}{\AA}{439}{559}
\refitem{Hartman, J.D., and Bakos, G.{\'A}.}{2016}{Astronomy and Computing}{17}{1}
\refitem{Kaluzny, J., \etal}{2013}{\AJ}{145}{43}
\refitem{Konacki, M., Torres, G., Jha, S., and Sasselov, D.D.}{2003}{Nature}{421}{507}
\refitem{Konacki, M., Torres, G., Sasselov, D.D., and Jha, S.}{2005}{\ApJ}{624}{372}
\refitem{Kooreman, C.J.}{1966}{Ann. Sterrew. Leiden}{22}{159}
\refitem{Kov{\'a}cs, G., Zucker, S., and Mazeh, T.}{2002}{\AA}{391}{369}
\refitem{Mennickent, R.E., Pietrzyñski, G., Diaz, M., and Gieren, W.}{2003}{\AA}{399}{L47}
\refitem{Mróz, P., \etal}{2015a}{\ApJS}{219}{26}
\refitem{Mróz, P., \etal}{2015b}{\Acta}{65}{313}
\refitem{Paczyñski B.}{1997}{~}{~}{in: Space Telescope Science Institute Series, ``The Extragalactic Distance Scale'', Ed. M. Livio (Cambridge Univ. Press), 273}
\refitem{Parenago, P.P.}{1931}{Peremennye Zvezdy}{3}{99}
\refitem{Pawlak, M., \etal}{2014}{\Acta}{64}{293}
\refitem{Pawlak, M., \etal}{2016}{\Acta}{66}{421}
\refitem{Pickering, E.C.}{1908}{Harv. Coll. Obs. Circ.}{137}{1}
\refitem{Pickering, E.C., and Leavitt, H.S.}{1904}{\ApJ}{20}{296}
\refitem{Pietrukowicz, P., \etal}{2013}{\Acta}{63}{115}
\refitem{Pietrzy{\'n}ski, G., \etal}{2012}{Nature}{484}{75}
\refitem{Pietrzy{\'n}ski, G., \etal}{2013}{Nature}{495}{76}
\refitem{Plaut, L.}{1948}{Ann. Sterrew. Leiden}{20}{3}
\refitem{Plaut, L.}{1958}{Ann. Sterrew. Leiden}{21}{217}
\refitem{Plaut, L.}{1971}{\AAS}{4}{75}
\refitem{Pojmañski, G., and Maciejewski, G.}{2004}{\Acta}{54}{153}
\refitem{Pojmañski, G., and Maciejewski, G.}{2005}{\Acta}{55}{97}
\refitem{Roberts, A.W.}{1895}{\AJ}{15}{100}
\refitem{Schwarzenberg-Czerny, A.}{1996}{\ApJ}{460}{L107}
\refitem{Soszyñski, I., \etal}{2011a}{\Acta}{61}{1}
\refitem{Soszyñski, I., \etal}{2011b}{\Acta}{61}{285}
\refitem{Soszyñski, I., \etal}{2013}{\Acta}{63}{21}
\refitem{Soszyñski, I., \etal}{2014}{\Acta}{64}{177}
\refitem{Soszyñski, I., and Udalski, A.}{2014}{\ApJ}{788}{13}
\refitem{Soszyñski, I., \etal}{2015}{\Acta}{65}{39}
\refitem{Swope, H.H.}{1938}{Ann. Harv. Col. Obs.}{90}{231}
\refitem{Szymañski, M., Kubiak, M., and Udalski, A.}{2001}{\Acta}{51}{259}
\refitem{Udalski, A., Kubiak, M., Szymañski, M., Ka³u¿ny, J., Mateo, M., and Krzemiñski, W.}{1994}{\Acta}{44}{317}
\refitem{Udalski, A., Szymañski, M., Ka³u¿ny, J., Kubiak, M., Mateo, M., and Krzemiñski, W.}{1995a}{\Acta}{45}{1}
\refitem{Udalski, A., Olech, A., Szymañski, M., Ka³u¿ny, J., Kubiak, M., Mateo, M., and Krzemiñski, W.}{1995b}{\Acta}{45}{433}
\refitem{Udalski, A., Olech, A., Szymañski, M., Ka³u¿ny, J., Kubiak, M., Krzemiñski, W., Mateo, M., and Stanek, K.Z.}{1996}{\Acta}{46}{51}
\refitem{Udalski, A., Olech, A., Szymañski, M., Ka³u¿ny, J., Kubiak, M., Mateo, M., Krzemiñski, W., and Stanek, K.Z.}{1997}{\Acta}{47}{1}
\refitem{Udalski, A., \etal}{2002a}{\Acta}{52}{1}
\refitem{Udalski, A., \etal}{2002b}{\Acta}{52}{115}
\refitem{Udalski, A., Szymañski, M.K., and Szymañski, G.}{2015}{\Acta}{65}{1}
\refitem{Wo¼niak, P.R.}{2000}{\Acta}{50}{421}
\refitem{Wo¼niak, P.R., Udalski, A., Szymañski, M., Kubiak, M., Pietrzyñski, G., Soszyñski, I., and ¯e\-bruñ,~K.}{2002}{\Acta}{52}{129}
\end{references}
\end{document}